\documentclass[aps,nofootinbib,
preprintnumbers]{revtex4-1}
\usepackage{latexsym}
\usepackage{graphics}
\usepackage{amsmath}
\usepackage{amsfonts}
\newcommand{\lp}{\left(}
\newcommand{\rp}{\right)}

\usepackage{amsmath}
\usepackage{amssymb}
\usepackage[paper=letterpaper,margin=1in]{geometry}
\usepackage{braket}
\usepackage{upgreek}
\newcommand{\abs}[1]{\ensuremath\left\lvert #1 \right\rvert}

\newcommand{\grad} {\nabla}
\newcommand{\lap} {\triangle}

\newcommand{\Ly}[1]{\lap_x #1 + \frac{a}{y} #1_y + #1_{yy}}

\newcommand{\R}{\mathbb R}

\newcommand{\beq}{\begin{eqnarray}}
\newcommand{\eeq}{\end{eqnarray}}

\newcommand{\dd}{\ensuremath\mathrm{d}}

\def\det{{\rm{det}}}

\usepackage{tikz}

\newtheorem{thm}{Theorem}[section]

\newtheorem{preremark}[thm]{Remark}

\numberwithin{equation}{section}

\begin{document}
\title{ Geodesically Complete Metrics and Boundary Non-locality in Holography:  Consequences for the Entanglement Entropy}
\author{Gabriele La Nave}
\affiliation{Department of Mathematics, University of Illinois,
Urbana, Il. 61820}
\author{Philip W. Phillips}
\affiliation{Department of Physics and Institute for Condensed Matter Theory,
University of Illinois, 1110 W. Green Street, Urbana, IL 61801}

\begin{abstract}

We show explicitly that the full structure of IIB string theory is needed to remove the non-localities that arise in boundary conformal theories that border hyperbolic spaces on AdS$_5$.  Specifically, using the Caffarelli/Silvestri\cite{caffarelli}, Graham/Zworski\cite{graham}, and Chang/Gonzalez\cite{chang:2010} extension theorems, we prove that the boundary operator conjugate to bulk p-forms with negative mass in geodesically complete metrics is inherently a non-local operator, specifically the fractional conformal Laplacian.  The non-locality, which arises even in compact spaces, applies to any degree p-form such as a gauge field.  We show that the boundary theory contains fractional derivatives of the longitudinal components of the gauge field if the gauge field in the bulk along the holographic direction acquires a mass via the Higgs mechanism. The non-locality is shown to vanish once the metric becomes incomplete, for example, either  1) asymptotically by adding N transversely stacked Dd-branes or 2) exactly by giving the boundary a brane structure and including a single transverse Dd-brane in the bulk.  The original Maldacena conjecture within IIB string theory corresponds to the former.  In either of these proposals, the location of the Dd-branes places an upper bound on the entanglement entropy because the minimal bulk surface in the AdS reduction is ill-defined at a brane interface.  Since the brane singularities can be circumvented in the full 10-dimensional spacetime, we conjecture that the true entanglement entropy must be computed from the minimal surface in 10-dimensions, which is of course not minimal in the AdS$_5$ reduction.  

\end{abstract}

\maketitle

\section{Introduction}

Intrinsic to Maldacena's conjecture \cite{maldacena} that supergravity (and string theory) on $d+1$-dimensional
 $\rm AdS$ space 
($\rm AdS_{d+1}$) times a compact manifold, a sphere in the maximally supersymmetric case, is equivalent to the 
large $N$ limit of $SU(N)$ conformal field theory in $d$ dimensions
 is the separation between bulk and boundary physics.  The impetus for such ideas originates from pioneering work of Susskind\cite{susskind}  and 't Hooft\cite{thooft}, who named the bulk-boundary correspondence in gravity {\it holography}.   However, as is well known\cite{Witten1998}, the boundary of any asymptotically AdS spacetime lives at infinity.  Hence, it does not inherit a well defined metric structure.  The structure it does acquire at the boundary is entirely conformal as can be seen from the Euclidean signature rendition
\beq
ds^2=\frac{dy^2+\sum_i d x_i^2}{y^2}
\eeq
of the  AdS metric.   The singularity at $y=0$ can be removed by considering the conformally equivalent metric $y^2ds^2$.  In fact, any metric of the form
\beq
\label{ew}
ds^2\rightarrow  e^{2w} ds^2
\eeq
would do the trick ($w$ a real function) , thereby laying plain the inherent conformal structure of the boundary.  

Hence, correlation functions of the conformal operators of the boundary theory should in principle encode the physics of quantum gravity in a spacetime that is asymptotically AdS.  Strictly speaking, however, the conformal field theory (CFT) only describes the physical excitations near the boundary. Precisely how far into the bulk this description\cite{KabatLL2011} applies remains an open question.  A key aspect of the mapping is that the CFT contains local operators.  Consider the example of a free field propagating in the bulk that obeys the Klein-Gordon equation.  The correspondence between the bulk and the boundary physics stems potentially from the equivalence between the partition functions 
\beq\label{corresp}
\langle e^{\int_{S^d}\phi_0{\cal O}}\rangle_{\rm CFT}=Z_S(\phi_0),
\eeq
in the two theories where ${\cal O}$ is the boundary operator, $\phi_0$ is the extension of the bulk field to the boundary and $Z_S$ is the supergravity partition function averaged over all double-pole metrics.  This form of the correspondence relies on an integration of the bulk action by parts and then an evaluation of the corresponding boundary terms\cite{gubser,Witten1998}.  Alternatively, an equivalence can be established by extrapolating the behavior of bulk correlators to the boundary\cite{bdhm}.  Near the boundary, the solutions scale asymptotically as
\beq
\phi(x,y)\approx y^{\Delta_-}\phi_0(x) + y^{\Delta_+} \mathcal O,
\eeq
where $y$ is the holographic coordinate (so that the conformal boundary is at $y=0$) and $\phi_0=\lim _{y\to 0} y^{\gamma -\frac{d}{2}} \phi(x,y)$, where $\gamma$ is a number such that $m^2= \frac{d^2}{4} - \gamma ^2$, and $ \Delta _{\pm} =\frac{d}{2} \pm\gamma$ .   The AdS/CFT duality\cite{bdhm,polchinski2010} in the extrapolation scheme dictates that we associate with this boundary behavior a corresponding  {\bf local} conformal operator ${\cal O}$ whose dimension is $\Delta$.  So, setting $\Delta =\Delta _+= \frac{d}{2} +\gamma$, the precise prescription\cite{polchinski2010} for accomplishing this is the limit
\beq
\label{operatoreq}
{\cal O}=C_{\cal O}\lim_{y\rightarrow 0} y^{-\Delta}\phi(x,y),
\eeq
where the boundary non-renormalizable term $y^{\Delta_-}\phi_0$ has been removed so that the limit is well defined.  One should think of as the BDHM\cite{bdhm} formulation of the AdS/CFT correspondence:
\beq
\label{BDMHeq}
\langle {\cal O}(x_1)\cdots  {\cal O}(x_n)\rangle_{CFT} =C_{\cal O}\lim_{y\rightarrow 0} y^{-n\Delta}\langle \phi(x_1,y) \cdots  \phi(x_n,y)\rangle_{bulk}.
\eeq


In this work, we show that when the bulk action, $Z_S(\phi)$, is a Gaussian theory, then for some values of the mass squared of the bulk field $\phi$, the operator $\mathcal O$ augmenting the boundary theory is an {\it anti-local} operator: the fractional Laplacian.  This is true regardless of the formulation that is used to express the bulk-boundary correspondence. We then argue -- following a logic reminiscent of the one adopted by Giddings (cf. \cite{gidding} and \cite{hs})--that since interactions turn off near the boundary, even in an interacting theory, the operator $\mathcal O$ must still be an anti-local operator (presumably having the fractional Laplacian as leading term).

Another way in which the interaction terms tend to vanish is for the $SU(N)$ theory with $N$ large. In the bulk interactive theory the bulk action can only be calculated by perturbative expansion, e.g. using Witten's graphs. Nonetheless, 
as shown in \cite{lmrs} and \cite{af}, when the interacting fields correspond to Kaluza-Klein modes of the compactified supergravity theory, say $\phi _i$, then, if we write the action as,
\beq S_{bulk}= \int d^{d+1} x \sqrt{-g} \left( \sum _i |\nabla \phi _i|^2 + m_i^2 \phi _i^2 +\sum _{i,j} \lambda _{ij}\, \phi _i^2\phi _j\right),
 \eeq
 then the coefficients $\lambda _{ij} = O(\frac{1}{N})$. This indicates that, as $N\to +\infty$, the operators $\mathcal O$ in the boundary which are dual to Kaluza-Klein modes must behave as the fractional Laplacian, for suitable values of the mass squared. 
 
The mathematics behind either the original or the BDHM\cite{bdhm} AdS/CFT correspondence is that of determining the asymptotic structure of solutions to the equations of motion, when approaching the conformal boundary. Three mathematical groups\cite{caffarelli,graham,chang:2010} have developed theorems that solve such boundary extension problems.  We first show that Eq. (\ref{operatoreq}) is explicitly of the form needed to apply the Caffarelli/Silvestre extension theorem.  Applying the theorem allows us to show that for a bulk field obeying the Klein-Gordon equation, Eq. (\ref{operatoreq}) is explicitly the fractional Laplacian acting on the boundary field $\phi_0$.  Unlike the normal Laplacian, the fractional Laplacian is explicitly a non-local operation in that it requires knowledge of the function everywhere for it to be evaluated.  Within the AdS/CFT conjecture as a whole\cite{maldacena,gubser,Witten1998,bdhm}, our work establishes a technical procedure for going between bulk fields defined by appropriate equations of motion and corresponding operators at the conformal boundary.    Although using the Caffarelli/Silvestre theorem requires that we equate the bulk field, namely the scalar field $\phi$ in $\mathbb H_d$ with $g=y^{\gamma-d/2}\phi$ in $\mathbb R^{d+1}$, the final result is more than a field redefinition.  This procedure results in a closed expression for the operator at the conformal boundary.  The difference with our result and the claim in the original conjecture is that  ${\cal O}$ is explicitly non-local.   The key results are summarized in Table 1.   We demonstrate that the non-locality is an intrinsic property of spacetimes that are geodesically complete.  We show explicitly that the Maldacena conjecture that local theories lie at the boundary of AdS spacetimes is recovered only if some degree of geodesic incompleteness is present in the bulk metric.  For example, stacking N branes transverse to the radial direction in the bulk leads to a local theory at the boundary in the asymptotic limit of $N\rightarrow\infty$.  As a result, our work implies that the strong form of the AdS/CFT duality with finite $N$ cannot hold without including non-local operators at the boundary.  Implicit in any form of the conjecture is the fixing of a vacuum in the boundary and the bulk.  While in general the two vacua might not be related, in our work the boundary vacuum emerges from the bulk.  This type of emergence is physical and explicit. In order to connect our formulation of the AdS/CFT correspondence with the more standard strong-weak formulation one must understand the interconnection between our vacuum and the one of N=4 SYM.  This will be addressed in a later publication.  Geodesic incompleteness poses a problem for the geometric interpretation\cite{rt2006} of the entanglement entropy as the minimal surface cannot cross a Dd-brane singularity.    We propose that a higher-dimensional geometric construction within the full type IIB string theory is necessary to retain the minimal surface idea.  Boundary non-localities appear as well for vector fields where the fractional exponent is governed by the mass of the bulk gauge field along the holographic direction.  This provides an explicit mechanism for producing anomalous dimensions\cite{anomdim} for boundary gauge fields. 

\begin{table}[h!]
\centering
\begin{tabular}{|c|c|}
\hline
{\rm Bulk Operator} & {\rm Boundary Operator: ${\cal O}=C_{\cal O}\lim_{z\rightarrow 0} y^{-\Delta}\phi(x,y)$ }\\
\hline
$|\nabla\phi|^2+m^2\phi^2$ & $(-\nabla)^\gamma\phi_0$\\
\hline
$F_{\mu\nu}F^{\mu\nu}+m^2A^2_y$ & $(-\nabla)^\gamma A_{\perp}$\\
\hline
\end{tabular}
\caption{Bulk-Boundary Correspondence on AdS$_{d+1}$ resulting from applying the Cafferelli/Silvestre extension theorem to ${\cal O}=C_{\cal O}\lim_{y\rightarrow 0} y^{-\Delta}\Phi(x,y)$ with $\Phi(x,y)$ the bulk field obeying the equation of motion specified in the Table. Here $\gamma=\sqrt{d^2+4m^2-p}/2$ and $A_\perp$ are the components of the gauge field perpendicular to the holographic coordinate, $y$.  Here $p$ is the degree of the form, which for a scalar is 0 and for the gauge field is $1$.}
\end{table}

\section{Preliminaries:  Evaluation of Eq. (\ref{operatoreq})}

To put our work in the context of the AdS/CFT correspondence, we review the standard procedure\cite{Witten1998} for a massive scalar field.  To this end, we work with the action
\beq
S_{\phi}=\frac12\int d^{d+1} u  \sqrt{g}\left(|\grad\phi|^2+m^2\phi^2 \right).
\eeq
 For the purposes of this initial discussion, we assume an AdS background with Euclidean signature (although we can and will work, {\it mutatis mutandis}, with the  general case of a black hole endowed with near horizon AdS geometry). The equations of motion  for the field $\phi$ are then simply given by,
\beq\label{eq:massive1}
-\Delta \phi+ m^2 \phi=0,
\eeq
 where $-\Delta =\grad _i\grad^i$ is the Laplacian.
It is a classical fact that this equation admits the existence of a unique solution on $\{(y,x_1, \cdots , x_d): y\geq 0\}$ with any given boundary value (the boundary is a sphere $S^d$, as described by  copies of $\R^d$ given by $y=0$ and the point $y=\infty$). 

As is standard\cite{Witten1998}, we assume that in the correspondence between $\rm AdS_{d+1}$ and conformal
field theory on the boundary, $\phi_0$  should
be considered to couple to a conformal
field ${\cal O}$, via: $\int_{S^d}\phi_0{\cal O}$.
Thus, in order to compute the two point function of ${\cal O}$, 
one must evaluate $S_\phi$ for a classical solution with boundary
value $\phi_0$. The equation of motion can be rewritten in the form,
\beq\label{eq:massive2}
-\Delta \phi -s (d-s) \phi=0,
\eeq
where $s$ is such that $m^2= -s (d-s)$ (i.e., $s= \frac{d}{2} + \frac{1}{2} \sqrt{ d^2+4m^2}$). Thus, as shown in Mazzeo and Melrose\cite{Mazzeo}, such a solution has the form,
\beq\label{asymptoticvalue}
\phi =F y ^{d-s} + G y^s,\quad F,G\in\mathcal C^\infty(\mathbb H),\quad F=\phi_0+ O(y^2), \quad G= g_0 + O(y^2),
\eeq
unless $s(d-s)$ belongs to the pure point spectrum of $-\Delta $. Here $\phi _0$ and $g_0$ are functions on the conformal boundary $\{ y=0\}$. A vast generalization of this fact, which we shall use later, can be found in \cite{Mazzeo}.
We refer to $\phi_0$ as the restriction of $\phi$ to the boundary of $\rm AdS_{d+1}$.
Operationally, the formal AdS/CFT correspondence can be established by  taking the finite part of the result of integrating $S_\phi$ by parts,
\beq \label{intbyparts}{\rm{pf} } \;
 \int_{ y >\epsilon}\;\left(\vert \partial \phi\vert ^2 -s(d-s)\phi^2\right)
dV_g 
=-d\int_{y=0} \; \phi _0\,g_0
\eeq
where $ {\rm{pf}} $ denotes the finite part of the divergent integral and $dV_g=\sqrt{g}d^{d+1}u$ is the volume form of $ds^2$. Therefore, $g_0$ must be the two-point function of $\mathcal O$.
We claim at this point that $g_0= G\mid _{y=0}$ is indeed the Riesz fractional Laplacian { \begin{footnote}{ The Riesz fractional Laplacian of a function $f$ defined on $\R^d$ is 
$(-\Delta)^\gamma f(x) = C_{d,s} \int_{\R^d} \frac{f(x) - f(\xi)}{\abs{x-\xi}^{d+2\gamma}} \dd \xi$
where $C_{d,s}$ is some normalization constant.}\end{footnote}} of $\phi_0$, $(-\Delta)^\gamma  \phi _0$, up to a constant factor.  

%
In order to show this, we need to appeal to a construction due to Caffarelli and Silvestre \cite{caffarelli}, which characterizes the Riesz fractional Laplacian $(-\Delta )^\gamma f$ of a function $f$ defined on $\R^d$ via an extension problem. Explicitly, what they showed is that given a function $f(x)$ defined on $\R^d$, a solution to
\begin{align}
g(x,0) &= f(x) \label{eq:dirichletboundary}\\
\Ly g &= 0. \label{eq:withy}
\end{align}
has the property that
\begin{equation}\label{caff-limit}  \lim _{y\to 0^+} y^a \frac{\partial g }{\partial y} =C_{d,\gamma}\; {(-\lap)^\gamma f} \end{equation}
for some (explicit) constant $C_{d,\gamma}$ only depending on $d$ and $\gamma = \frac{1-a}{2}$.

Now we observe that if $\phi$ solves the massive problem \eqref{eq:massive1} (in fact its representation in the form of Eq. \eqref{eq:massive2}), then an easy computation shows that the function
\beq
 g=y^{\gamma -\frac{d}{2}}\, \phi, \qquad \gamma:=\frac{ \sqrt{d^2+4m^2}}{2} 
 \eeq
 solves the Caffarelli-Silvestri extension problem, Eqs. (\eqref{eq:dirichletboundary}) and (\eqref{eq:withy}). But since a solution, $\phi$, to the massive problem has the asymptotic expansion (using that $s= \frac{d}{2}+\gamma$),
 \beq
\phi =F y ^{\frac{d}{2}-\gamma} + G y^{\frac{d}{2}+\gamma},\quad F,G\in\mathcal C^\infty(\mathbb H),\quad F=\phi_0+ O(y^2), \quad G= \mathcal O + O(y^2),
\eeq
it then follows that
\beq
g=y^{\gamma -\frac{d}{2}}\, \phi =F  + G y^{2\gamma},\quad F,G\in\mathcal C^\infty(\mathbb H),\quad F=\phi_0+ O(y^2), \quad G= g_0 + O(y^2).
\eeq
 Now we make two observations. On the one hand by the asymptotic expansion of $g$ above, it must be that
 \beq	\label{CSex}
 \lim_{y\to 0} y^{1-2\gamma}\frac{\partial g}{\partial y}= 2\gamma g_0.
 \eeq
 On the other, by the result of Caffarelli and Silvestri\cite{caffarelli} this limit is $(-\lap)^s \phi _0$, up to a constant factor, thus showing that the two point function of the operator $\mathcal O=(-\Delta)^\gamma \phi_0$ is a multiple of $\vert x-x'\vert ^{-d -2\gamma}$.  
 
To address the BDHM formulation\cite{bdhm}, we note that the Cafferelli/Silvestre extension equation, Eq. (\ref{CSex}), is precisely of the form of the operator identity in Eq. (\ref{operatoreq}).  To make this more transparent, we note that powers of $y$ can be traded for derivatives with respect to $y$ since Eq. (\ref{CSex}) is based on the asymptotic expansion of the solutions for the equations of motion, namely Eq. (\ref{asymptoticvalue}).  Consequently, we rewrite Eq. (\ref{operatoreq}) as
 \beq
 {\cal O}=C_{\cal O}\lim_{y\rightarrow 0} y^{1-2\gamma}\partial_y\phi(x,y).
 \eeq
 With the substitution $g=y^{\gamma-d/2}\phi$, this equation is precisely of the form of Eq. (\ref{CSex}) thereby offering another proof that the fractional Laplacian is the operator dual of the bulk free scalar field.  We note that the fractional Laplacian in flat space is a conformal operator and in fact this is a general feature of operators obtained in this fashion via a scattering process (see  Eq. IV.20). This should be kept in mind as one considers the true boundary of AdS, which is a sphere in Euclidean signature.
 
 It is remarkable that one attains all non-negative real values of $\gamma=\frac{ \sqrt{d^2+4m^2}}{2} $, even in this non-conformal picture, as $m^2\geq -\frac{d^2}{4}$ is allowed by the Breitenlohner-Freedman (BF)\cite{BF} bound for stability.  However, the theory thus presented has the unfortunate feature of being non-local however.  As we will see such non-locality is unavoidable and present even in the conformal construction. One should observe that the negative BF bound is of course only possible for states $\phi$ such that $\int _{\mathbb H} \frac{ dydx}{y^d}\, \phi^2$ is not finite (otherwise the  mass term would have to be positive).
 In fact, in complete generality, J. Lee (\cite{lee}) proved that the essential spectrum of any asymptotically Einstein manifold is bounded from below by $-\frac{d^2}{4}$ and in the non-compact case,  there are no embedded eigenvalues, thus ensuring that there are no square-normalizable (i.e., renormalizable) eigenfunctions.

 \section{Stability and Planck length}

Of course, the stability condition depends on the AdS radius of curvature, $L$.  To  introduce this length, we rescale the metric,
\beq
d\tau^2=L^2\frac{dy^2+\eta _{\mu \nu} d x^\mu dx^\nu}{y^2},
\eeq
accordingly.
Using this Lorentzian metric, we write the action for the Klein-Gordon field as
\beq
\begin{aligned}
S&= -\frac{1}{2L^{d-1}}\int dy d^d x \sqrt{g}\left( g^{\mu \nu} \partial _\mu \phi \partial _\nu \phi + m^2\phi ^2\right)\\&= -\frac{1}{2} \int  \frac{ dy d^d x}{L^{d+1}} \; \left( y^2 \partial _y \phi  \partial _y \phi  + y^2 \eta ^{\mu \nu} \partial _\mu \phi \partial _\nu \phi + m^2 L^2 \phi ^2\right).\end{aligned}
\eeq
Therefore, the relevant equation to establish the asymptotic structure of Eq. \eqref{asymptoticvalue}, is\beq
\gamma= \frac{1}{2} \sqrt{d^2 + m^2 L^2} .
\eeq
Performing the change of variables\cite{polchinski2} $z=\ln y$ and setting $\phi = y^\frac{d}{2} \psi$, we obtain
\beq
S= -\frac{1}{2}\int dz d^d x  \left( \partial _z \psi  \partial _z \psi  + e^{-2z} \eta ^{\mu \nu} \partial _\mu \psi \partial _\nu \psi +[ m^2 L^2+\frac{d^2}{4} ] \psi ^2\right)\
\eeq
which shows that the Hamiltonian is a sum of squares (up to adding the boundary term $-\frac{d^2}{4}\int   d^d x \psi ^2\mid _{z=-\infty} ^{z=+\infty}$) provided that
\beq
m^2 L^2\geq -\frac{d^2}{4}
\eeq
which is the BF bound.
This of course still makes $\gamma$ possibly arbitrarily close to $0$, but as $L$ grows, this occurs with a mass terms $m^2$ which are increasingly close to $0$,
\beq
\lim _{L\to +\infty}  -\frac{d^2}{4 L^2}=0.
\eeq
Therefore, the non local phenomenon present at the boundary theory requires the presence of less strange (tachyonic) matter in the bulk as the black hole radius increases.

 \section{Conformal Holography}
 
 Away from flat space, the fractional Laplacian is not a conformal operator.  To ensure conformality, we must include a conformal sector in the starting action.   Namely, we consider the following action in the bulk
 \beq
 S=S_{gr}[g]+ S_{\rm matter}(\phi),
 \eeq
where the standard Einstein-Hilbert action (with the Gibbons-Hawking 
boundary term) is given by
\beq
S_{gr}[g]=-\frac{1}{2\kappa^2}\left[\int_M d^{d+1}x\sqrt{g}R+
	\int_{\partial M}d^dx\sqrt{h}2K\right],
\eeq
$\kappa^2\equiv 8\pi G_{d+1}$, $h$ is the induced metric
on $\partial M$ and  $K$ is the trace of the extrinsic curvature of
 the boundary. The new term is something we name {\it conformal matter} given by the action
\beq
S_{\rm matter}=\int_M d^{d+1}x\sqrt{g}\mathcal{L}_m,
\eeq
with 
\beq\label{Lagr-confmatter}
\mathcal{L}_m:= \vert \partial \phi \vert ^2 + \left(m^2+\frac{d-1}{4d}R(g)\right) \phi ^2.
\eeq
The new term in $\mathcal{L}_m$,  $R(g)\, \phi ^2$, contributes to the Euler-Lagrange equations in the form of the {\it conformal Box} operator,
\beq
\Box_g^{conf} \phi= \frac{1}{\sqrt{g}} \partial_\mu (\sqrt{g} g^{\mu \nu} \partial_\nu \phi)-\frac{d-1}{4d}R_g\phi= \Box_g \phi -\frac{d-1}{4d}R_g\phi.
\eeq

The advantage of using a conformal action (as part of the total action) is that one incorporates the fact that the boundary only has a well defined conformal class of metrics (arising from conformally compactifying AdS) into the theory. The boundary theory operators $\mathcal O$ naturally correspond to {\it conformal Laplacians}. Moreover, in the case of a conformal Einstein manifold (such as the hyperbolic space), simplifications arise. Recall that on a Riemannian manifold $(M,g)$ of dimension $N=d+1$, the conformal Laplacian is
\beq
L_g\equiv-\Delta_g+\frac{N-2}{4(N-1)}R_g=-\Delta_g+\frac{d-1}{4d}R_g,
\eeq
which, after a conformal change of metric, $\hat g=e^w \,g$, transforms as
\beq\label{conformal-laplacian1}
L_g(\psi)=(e^w)^{\frac{d+3}{2}}L_{\hat g}\lp (e^w)^{-\frac{d-1}{2}}\psi\rp.
\eeq
For the hyperbolic metric $g= \frac{ dy^2+\sum _{i=1}^d\, dx_i^2}{y^2}$, the scalar curvature is $R_{g_{\mathbb H}}=-d(d+1)$, so that
\beq
\label{conformal-laplacian2}
L_{g_{\mathbb H}}=-\Delta_{g_{\mathbb H}}-\tfrac{d^2-1}{4}
\eeq
and now, the BF\cite{BF} stability bound becomes $m^2\geq -\frac{1}{4}$.  This condition actually is independent of the dimensionality because we can write $m^2-\tfrac{d^2-1}{4}= -s (d-s)$ with $s= \frac{d}{2} + \frac{\sqrt{ 4m^2+1}}{2}$ which is equivalent to $ \gamma:=\sqrt{ 4m^2+1}\geq0$.
The conformal dimension of the field $\mathcal O$ is exactly $d+\gamma$.

In complete  generality, one defines an asymptotically $d+1$ AdS space-time as a $(d+1)$-dimensional space time $(M, d\tau ^2=g^+ _{\mu \nu}\, dx^\mu \otimes dx^\nu)$ such that $M$ has a topological boundary $X$  characterized as follows:
\begin{enumerate}
\item There exists a function $\rho>0$ in $M$ such that $\rho \mid_X=0$ and $\nabla \rho \neq 0$ (i.e. $\rho =0$ is a {\it defining} function for the boundary,
\item $\rho ^2 \, d\tau ^2\mid _X$ is a smooth Lorentzian metric,
\item (The space looks like AdS at infinity) there exists a diffeomorphism $\Psi : \{ 0<\rho < \rho _0\} \to \{0<y < y _0\} $ and real numbers $\rho _0, y_0>0$ (here $y$ is the coordinate/defining function of the boundary on AdS) such that $d\tau ^2= \Psi^* \left( \frac{dy^2 +\eta _{\mu \nu} dx^\mu \otimes dx^\nu}{y^2}\right) + O(\rho ^2)$ for $\rho >\rho _0$, and 
\item $d\tau ^2$ satisfies the Einstein equations: $R_{\mu \nu} -\frac{1}{2} R g_{\mu \nu} + \Lambda g_{\mu \nu } = 8\pi T_{\mu \nu} $.
\end{enumerate}

In this context, we still propose an AdS/CFT type correspondence, but with the Lagrangian given by the conformal matter equation above, namely Eq. (\eqref{Lagr-confmatter}).
Now the correspondence requires that we find solutions to the classical equations of motion,
\beq\label{confeqmot}-\Delta_g\phi +\frac{d-1}{4d}R_g\phi = m^2\phi
\eeq
and then perform the same scattering process we useded earlier in the classical theory. In general, due to the presence of the potential part 
$\frac{d-1}{4d}R_g\phi$, this analysis is considerably more complicated than the one performed in the classical case for AdS, and it tends to be very different even from the classical case of asymptotic AdS gauge/gravity duality (where one merely studies classical solutions of motion: $-\Delta_g\phi  = m^2\phi$)

Nonetheless, this theory becomes considerably easier in the case that $T_{\mu \nu}=0$. In this case, again switching to Euclidean signature, we can infer from the Einstein equation that the scalar curvature $R(g)$ has to be constant which we normalize such that,$R_{g_{\mathbb H}}=-d(d+1)$.
In this circumstance, the classical equations of motion for conformal matter (i.e., Eq. \eqref{confeqmot}) reduce to,
\beq\label{eq:massiveconfmatterTzero}
-\Delta \phi+\left( m^2-\tfrac{d^2-1}{4}\right) \phi=0.
\eeq

We write yet again this equation in the form,
\beq\label{eq:massive3}
-\Delta \phi -s (d-s) \phi=0,
\eeq
where $s$ is such that $ m^2-\tfrac{d^2-1}{4}= -s (d-s)$ (i.e., $s= \frac{d}{2} + \frac{1}{2} \sqrt{ 4m^2+1}$). Thus, setting $\gamma:= \frac{1}{2} \sqrt{ 4m^2+1}$, such a solution has the form,
\beq
\phi =F y ^{\frac{d}{2} -\gamma} + G y^{\frac{d}{2} +\gamma} ,\quad F,G\in\mathcal C^\infty(\mathbb H),\quad F=\phi_0+ O(y^2), \quad G= g_0 + O(y^2).
\eeq
 Here $\phi _0$ and $g_0$ are functions on the conformal boundary $\{ y=0\}$. Next, as in the case of the classical Laplacian (as opposed to the conformal one we are analyzing here), if we set
 \beq \label{gamma-conf}
 g=y^{\gamma -\frac{d}{2}}\, \phi, \qquad \gamma:=\frac{ \sqrt{4m^2+1}}{2},
 \eeq
 one readily finds that $g$ solves the Caffarelli-Silvestri extension problem, Eqs. \eqref{eq:dirichletboundary} and \eqref{eq:withy}. 
 It is now plain that,
 \beq
g=y^{\gamma -\frac{d}{2}}\, \phi =F  + G y^{2\gamma},\quad F,G\in\mathcal C^\infty(\mathbb H),\quad F=\phi_0+ O(y^2), \quad G= g_0 + O(y^2)
\eeq
and that by the asymptotic expansion of $g$ above,
 \beq
 \lim_{y\to 0} y^{1-2\gamma}\frac{\partial g}{\partial y}= 2\gamma g_0.
 \eeq
 
In the general case, we consider the asymptotic solutions to Eq. \eqref{confeqmot} and define the scattering operator as follows.
 Solutions to
\begin{equation}\label{equation-GZ}
-\Delta_{g} u-s(d-s)u=0,\quad\mbox{in } X
\end{equation}
have the form
\begin{equation}\label{general-solution}u = F \rho ^{d-s} + W\rho^s,\quad F,W\in\mathcal C^\infty(X),\quad F|_{\rho=0}=f,\end{equation}
for all $s\in\mathbb C$ unless $s(d-s)$ belongs to the pure point spectrum of $-\Delta_{g}$.
The {\it scattering operator} on $M$ is defined as $S(s)f = W|_M$.

Following \cite{chang:2010},  we define the conformally covariant fractional powers of the Laplacian (on the conformal boundary) as
 \begin{equation}\label{P-operator} P_\gamma[d\tau ^2, h] := D_\gamma \; S\lp\frac{d}{2}+\gamma\rp,\quad D_\gamma=2^{2\gamma}\frac{\Gamma(\gamma)}{\Gamma(-\gamma)}.\end{equation}
for $s=\frac{d}{2}+\gamma$, $\gamma\in \lp 0,\frac{d}{2}\rp$, $\gamma\not\in \mathbb N$. 
%
One readily sees that $P_\gamma\in  (-\Delta_{\hat g})^\gamma+\Psi_{\gamma-1}$, where  $\Psi_{\gamma-1}$ is a pseudo-differential operator of order $\gamma -1$.

By the property of $S$ proven in \cite{graham}, one has,
\begin{equation}\label{confomral-invariance}P_\gamma[d\tau ^2,h_v]\phi = v^{-\frac{d+2\gamma}{d-2\gamma}} P_\gamma[d\tau ^2, h] \lp v\phi\rp,\end{equation}
where,
\begin{equation}\label{change-metric}h_{v}=v^{\frac{4}{d-2\gamma}}h,\quad v>0,\end{equation}
and $[h]$ is the conformal infinity and 
\begin{equation}\label{scattering}P_\gamma f=d_\gamma S\lp\tfrac{d}{2}+\gamma\rp=d_\gamma\;h.\end{equation}
In this context, the operator $\mathcal O$ is found using,
\begin{equation}
{\rm{pf} } \;
 \int_{ y >\epsilon}[\vert \partial \phi \vert ^2-\left( s(d-s)+\frac{d-1}{4d}R(g)\right)\phi ^2]
dV_g 
=-d\int_{\partial X}dV_h  f\, P_\gamma[d\tau^2,h] f. 
\end{equation}
Consequently, the corresponding boundary operator is $P_\gamma$ which persists under any change to the bulk metric as long as the conformal boundary remains unchanged.  

\subsection{ A few words on the conformal Laplacian}

A choice of a Lorenzian metric {\begin{footnote} {We could discuss here independently of the signature, but we choose, for definiteness of notation, to use a Lorenzian structure } \end{footnote}} on a manifold $M$ of dimension $d+1$ is equivalent to a choice of an {\it orthonormal } frame bundle of $T^*M$. Choosing a conformal class is equivalent to a reduction of the structure group.
For any real number $\alpha\in \mathbb R$, one obtains a ~1-dimensional irreducible representation of $SO(1,d)$ given by $\det (A) ^{\frac{\alpha}{d+1}}$ which gives rise to a line bundle $L_\alpha$ for a fixed conformal structure. Choosing a metric $g$ in the conformal class $[g]$ is tantamount to choosing a trivialization $\tau _{g,\alpha} : L^\alpha \to M\times \mathbb R$ and changing $g$ by $e^{2w}\, g$ has the effect of changing the trivialization to  $e^{-\alpha w}\tau_{g,\alpha}$. The proper way of formulating the conformal box operator is to think of it as an operator,
\beq\Box_g^{conf} : L^{\frac{d-1}{2} } \to L^{\frac{d}{2} +1},
 \eeq
 that connects two vector bundles.
Given a vector bundle $E$ endowed with a Hermitian connection $\nabla_E$, more generally one defines the conformal Laplacian as
\beq
\Delta _E^{conf}\equiv\nabla_E ^* \nabla_E + \frac{ d-1}{4d} \, R,
\eeq
where
\beq
\nabla ^*_E \nabla_E = - \frac{1} {|g|^{\frac{1}{2}}}\, (\nabla_E) _\mu \left(g^{\mu \nu} |g|^{\frac{1}{2}} (\nabla_E) _\nu\right)
\eeq
and after suitably trivializing sections of $E\otimes L^w$ we can think of it as an operator with
\beq
\Delta _E^{conf} : \Gamma (E\otimes L^w) \to  \Gamma (E\otimes L^{w+2}).
\eeq
In this paper, we will suppress the line bundles, $L^\alpha$, by fixing a conformal representative.

%

\section{Gauge Theory}

In order to follow Witten\cite{Witten1998} closely, we again switch to Euclidean signature.
Here we consider adding the Lagrangian,
\beq
\mathcal L _G:= \frac{1}{2} \int _{AdS} \; F\wedge \* F,
\eeq
where $F=dA$ is the filed strength of the ~1-form $A$. The classical equations of motion are then (equivalent to) Maxwell's,
\beq d (\star dA)= 0,
\eeq
and in fact, the previous scattering process can be repeated, {\it mutatis mutandis}, as follows.
Given a ~1-from $\sum a_i dx_i$ on the conformal boundary, we want to solve for solutions to 
\begin{align}
 &d (\star dA)= 0\\ & A\mid _{y=0} = \sum a_i dx_i.
\end{align}
It is a standard consequence of the Weitzenb\"ock formula, which relates the Hodge Laplacian to the standard Laplacian, that the previous equation is related to 
\beq
\Delta A_\mu-dA_\mu= 0\quad A_\mu|_{y=0}=a_\mu\,\mu\neq y.
\eeq

\subsection{Higgs Mechanism for Fractional Gauge Fields}

In this section we describe how the process of symmetry breaking along the holographic direction gives rise to fractional Laplacians acting on Gauge fields at the boundary.
We describe here for simplicity just the case in which the gauge group is $U(1)$. We consider the Lagrangian,
 \beq
  \mathcal L = D_\mu \phi ^* D^\mu \phi - m^2 \phi ^* \phi - \lambda (\phi ^*\phi )^2 - \frac{1}{4} F_{\mu \nu }F^{\mu \nu},
 \eeq
 where $\phi$ is a function only of the radial coordinate, $y$.
 This Lagrangian is  invariant not only under the $U(1)$ transformation but also the complexified gauge group $\mathbb{C}^*$\cite{witten1991}.  For generality, we consider this larger gauge group here as it generates negative masses of the gauge field.  Hence, we consider a transformation of the form,
 \beq
 \phi \to e^{-i\theta(y,x)}\phi,
 \eeq
 with $A_\mu \to A_\mu -\frac{1}{e} \partial _\mu \theta$, where $\theta$ can be complex.
As is standard, we expand around the  vacuum expectation,
\beq
\langle \phi \rangle _0=\frac{v}{\sqrt{2}}.
\eeq
In other words, we break the $U(1)$ symmetry in the radial direction by writing,
\beq
\phi = e^{i\frac{\xi}{v}}\; \frac{v+\psi}{\sqrt{2}},
\eeq
where $\xi=\xi(y)$ is merely a function of the holographic direction.
Then the standard symmetry breaking,
\beq
\begin{aligned}\label{RGu1break}
&\hat \phi=  e^{-i\frac{\xi}{v}}\phi = \frac{v+\psi}{\sqrt{2}}\\
& \hat A_\mu = \to A_\mu -\frac{1}{e} \partial _\mu \xi,
\end{aligned}
\eeq
produces the Lagrangian
 \beq
 \begin{aligned}
  \mathcal L &=\frac{1}{2} \partial_\mu \psi ^* \partial ^\mu \psi - m^2 \phi ^* \phi  - \frac{1}{2} \psi ^2 (3\lambda v^2+m^2)+ \lambda v \psi ^3- \frac{1}{4}\lambda \psi ^4\\&- \frac{1}{4} \hat F_{\mu \nu }\hat F^{\mu \nu}+ \frac{1}{2} e^2 v^2  \hat A_\mu  \hat A^\mu + \frac{1}{2} e^2 v^2  (\hat A_\mu )^2 \psi (2v+\psi).
  \end{aligned}
 \eeq
 We can now apply the previous analysis to obtain terms of the kind $(-\Delta )^\gamma a_\mu$ at the boundary, where $\gamma=\sqrt{(ev)^2+d^2-1}$. 
 
 Observe that Eq. \eqref{RGu1break} shows that the $\hat A _\mu= A_\mu$ for $\mu \neq 0$ (i.e. in the non-holographic directions). Therefore we have clearly broken the symmetry merely in the holographic direction, thus leaving the boundary theory free to have any type of symmetry we please. 
Consequently,  we have provided a mechanism for understanding how boundary theories proposed recently\cite{gabadadze,anomdim} acquire gauge fields with fractional dimensions.  Results for the boundary form of the operators is summarised in Table 1.

\section{ Branes in Action: Maldacena's Duality on Incomplete Metrics}

How do we then recover Maldacena's conjecture that local conformal theories lie at the boundary of AdS spacetimes?  A crucial detail in the Maldacena\cite{maldacena} construction based on type IIB string theory is the $N$  D3 branes which he stacked transversely in the bulk.  We show explicitly here that it is only from these branes in the asymptotic limit that the gauge-gravity correspondence is free of non-local interactions.  That is, only when such branes are retained does the conformal theory field theory on the boundary have explicitly local operators.  

Recall from Horowitz and Strominger\cite{pbranes} that there is a black brane solution of IIB string theory which is spherically symmetric. Part of the low energy action from string theory is given explicitly\cite{pbranes} by
\beq
S= \int d^{10} x \sqrt{-g} \left( e^{-2\phi} (R+ 4 |\nabla \phi |^2 ) - \frac{ 2 e^{2\alpha \phi}}{(D-2)\!} F^2\right),
\eeq
where $F$ is a closed $D-2$-form. We take $D=7$ and the extremal solution (with no event horizon) is given by
\beq
\begin{aligned}
&ds^2 _L= H^{-1/2}(r) \, \eta _{\mu \nu} dx^\mu dx^\nu+  H^{1/2}(r) \, \delta _{mn} dx^m dx^n\\
&H= 1+\frac{L^4}{r^4}, \;\; L^4 =4\pi gN\alpha '^2, \; r^2= \delta _{mn} x^m x^n,
\end{aligned}
\eeq
where $N$ here is the number of stacked $D3$-branes (or flux of the black hole), $\alpha'$ is the string tension and $g$ the coupling constant. We now observe that the $L$ appears as a rescaling of the AdS metric 
and the rescaling property (Eq. \eqref{conformal-laplacian1}), yields
\beq
\Box _{ds^2}^{conf}(\phi)+m^2 \phi=L^2 \lp \Box_{ds^2_L}^{conf} + \frac{m^2}{L^2}\rp \phi
\eeq
whence we derive that the equations of motion in the $ds^2_L$ metric are equivalent to
\beq 
\lp \Box_{ds^2_L}^{conf} +\frac{m^2}{L^2}\rp \phi=0
\eeq
with $m^2$ the mass-squared in the $L=1$ theory (bounded from below by the BF bound\cite{BF})
thus showing that the boundary fractional Laplacians are of the type $(-\Delta )^\gamma$ with $\gamma = \frac{ \sqrt{4\frac{m^2}{L^2}+1}}{2}$. Since $\lim _{L\to +\infty} \gamma=\frac{1}{2}$, this shows that strictly as $L\to +\infty$, the non-localities disappear.  This proves our assertion that a conformal theory with purely local operators obtains only in the limit of an infinite number of transversely stacked branes.  

\begin{figure}
	\includegraphics[scale=0.5]{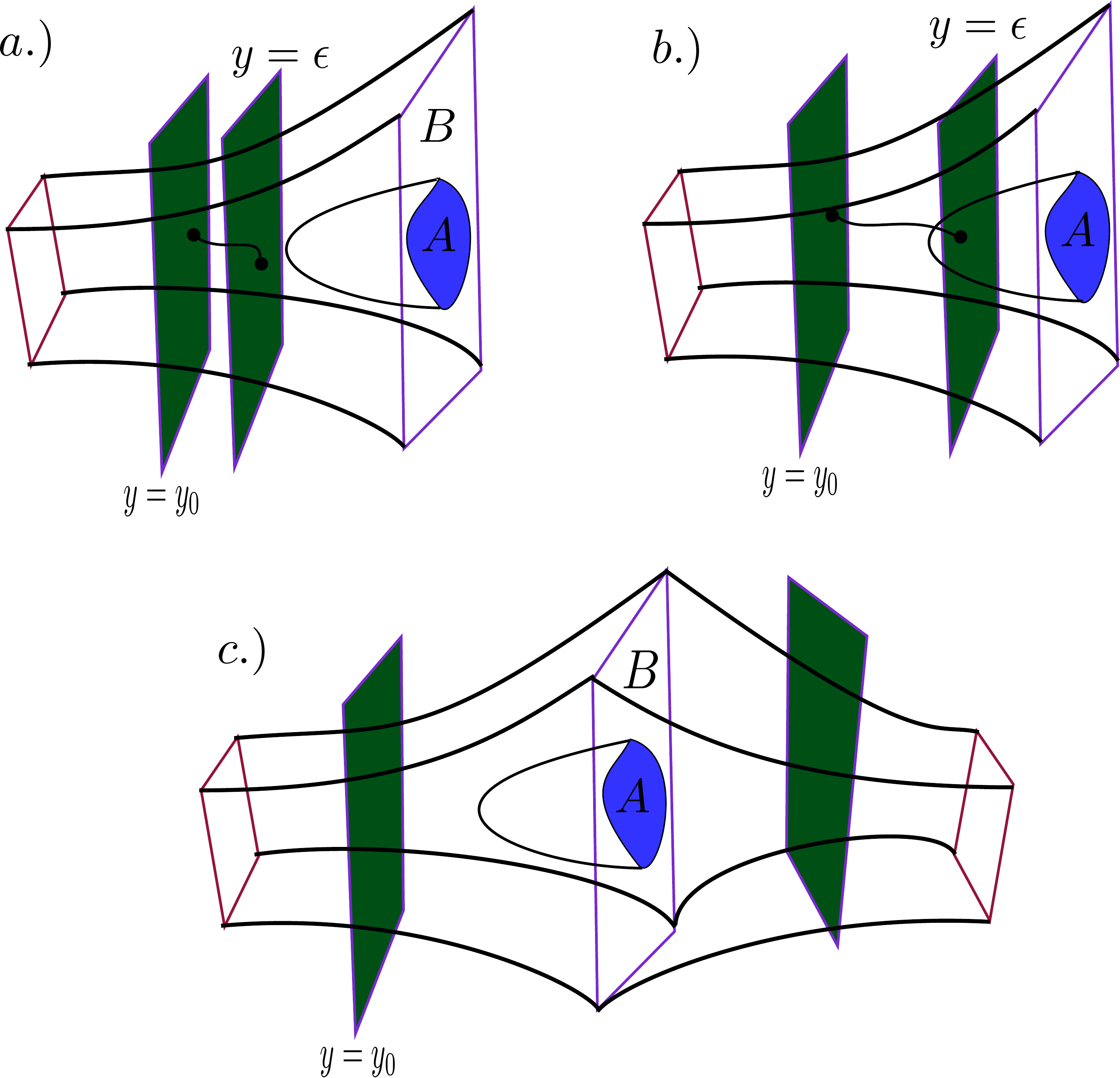}
    \caption{Hyperbolic space with two D3-branes:  a.) The two two branes lies outside the minimal surface used to compute the entanglement entropy between regions A and B, b) the physical impossibility of one of the D3-branes lying within the minimal surface and c) the limiting case in which $\epsilon\rightarrow 0$.  In the latter case, the metric doubles, the boundary vanishes as as a result so does the entanglement entropy.}
    \label{fig1}
\end{figure}

Alternatively, consider the string IIB solution whose background metric we write in general form as
\beq\label{iib-metric}
ds^2=f^{-1/2}\eta_{\mu\nu} dx^\mu dx^\nu+f^{1/2}\delta_{mn} dx^m dx^n,
\eeq
where $\mu, \nu =0,1,2,3$ and $m, n= 4, 5, 6, 7, 8,9$. The metric is on $\mathbb R^{3,1} \times K_6$ for some Einstein ~6-manifold $K_6$. 
The equations of motion dictate for $f$ to be a function of the transverse coordinates satisfying
\beq 
\Delta f = (2\pi )^4\, \alpha '^2 g  \, \rho,
\eeq
where $\rho= \rho (x^4, \cdots, x^9)$ is the density of Dd-branes. For instance, the standard solution is obtained by choosing $f=H$ with $r^2= x_m x^m$ as above, so that $\rho$ is a delta function counted with multiplicity determined by $L$ (hence the description of it as a stacking of $N$ branes positioned at the "horizon" $r=0$). 
 In this application we take $f$ to be a harmonic function that has a brane singularity at $r=\epsilon$ and another transverse brane somewhere in the bulk at $r=r_0$ (these are strictly speaking walls as they are co-dimension 1).  We are interested in the limit in which the D-brane approaches the boundary as illustrated in Fig. (\ref{fig1}); that is, $\epsilon\rightarrow 0$.  
 It is clear from the description of the singularity of the Laplacian of $f$ that near the singularity, $f$ is an absolute value singularity. It is then easy to construct solutions of this type that exhibit a full $\mathbb Z _2$ symmetry in the limiting configuration.
 
 We can make this supergravity argument come to light in a simple example of the Randall-Sandrum\cite{lrandall} type of metric in which the absolute value singularity is explicitly manifest. Our argument  works perfectly well in the IIB supergravity model, but for the sake of expository clarity we present this simpler model instead. We consider the ~5-dimensional spacetime with $ds^2= - e^{-2|y|/L} g_{\mu \nu} dx^\mu dx^\nu + dy^2$, which we think of as a fluctuation of the $3+1$ directions of the Randall-Sundrum metric $ - e^{-2|y|/L} \eta_{\mu \nu} dx^\mu dx^\nu + dy^2$ where $L$ is a length scale depending only on the mass $M_5$ (the analogue of the Planck mass) and the (negative) cosmological constant.   The presence of $|y|$ in the exponential guarantees that the metric is geodesically incomplete.  Such incompleteness has no affect on the connectedness of the boundary as guaranteed by the Witten-Yau theorem\cite{wittenyau,reidwang2012}.   As in Randall-Sundrum\cite{lrandall} we consider the $y$ direction to take values in the  quotient of the circle $S^1/{\mathbb Z_2}$ (which we think of as the interval $[-\pi R, \pi R]$ with the points $y$ and $-y$ identified). 
Then, since the coefficients of the metric at $y=\pi R$ are $e^{-2\pi R/L}g_{\mu \nu}$, the effective action of a massive particle at the brane positioned at $y=\pi R$ is proportional to
\beq 
\begin{aligned}
&\int d^4x \sqrt{-g} e^{-2\pi R/L}\left( e^{2\pi R/L}g^{\mu \nu}\partial_\mu \phi \partial_\nu \phi +m^2\phi ^2\right)\\&= \int d^4x \sqrt{-g}\left( g^{\mu \nu}\partial_\mu \hat \phi \partial_\nu \hat \phi +m^2 e^{-2\pi R/L}\hat \phi^2\right),
\end{aligned}
\eeq
where $\hat \phi = e^{-\pi R/L}\phi$. This clearly shows that for $R/L$ sufficiently large, the negative (effective) mass terms $m^2 e^{-2\pi R/L}$ again become asymptotically positive, thereby leading to a vanishing of the scalar solutions which give rise to the non-locality. The largeness of $R/L$ is of course an indication of a wall singularity which causes a "warping" of the compact manifold, in the language of \cite{cpv}.  We see explicitly then that incompleteness coupled with a wall singularity are needed to rid the boundary theory of non-locality. 

 This argument can be generalized beyond the Randall-Sundrum metric. In hyperbolic space, the mass of a string joining the two branes grows quadratically as $(\ln \epsilon)^2$.  Once  $|\ln\epsilon|>2\pi\sqrt{\alpha'}$, the mass becomes positive\cite{zwiebach}.  It is this mass that sets the scale for the masses of bulk scalar fields.  The solutions to the scalar field equations of motion we found earlier which give rise to the non-local boundary interactions are no longer valid should $\epsilon$ be sufficiently small so that the mass is positive, that is, a violation of the BF bound\cite{BF}.  Hence, any type of Dd-brane placed transverse to the holographic direction in a geometry in which the boundary is viewed as a brane singularity is sufficient to kill the non-local interactions found here.  The essence of this argument is that transverse walls break the completeness of the metric in the holographic direction.  Once this completeness is broken, locality of the boundary theory obtains.

%
%
 
\section{Entanglement}

D-brane bulk singularities also affect the geometric interpretation of the entanglement entropy\cite{rt2006}.   Computing the entanglement entropy of two regions in the boundary separated by a region $\Omega_A$ simply requires delineating the bulk minimal surface on AdS$_5$ that has $\Omega_A$ as its Dirichlet boundary condition.  Any such surface cannot remain minimal if it traverses a singularity in the bulk (see Fig. (\ref{fig1}b)), such as a D-brane. In the construction in Fig. (\ref{fig1}), if the D3-brane located at $y=\epsilon$ lies outside the minimal surface, the geometric interpretation of the entanglement entropy remains unaffected.  However, as $\epsilon$ approaches the boundary, the minimal surface has to shrink to avoid the D3-brane, thereby leading to a vanishing of the entanglement entropy in the limit $\epsilon\rightarrow 0$.   The singularity that arises in this limit depends on the type of D3-branes that are in the 5-dimensional theory.  If the D-brane arises from a reduction of a D3-brane in the 10-dimensional theory, then the brane stacking problem of Maldacena\cite{maldacena} arises, which we treated previously.  However, should the D3-brane arise from a D9-brane as in the previous section, then a wall singularity arises at the boundary resulting in a doubling of the metric.  In this case, the  metric resembles that of Randall-Sundrum\cite{lrandall} and, as a result, is incomplete.  Interestingly, only in the non-compact limit, $R/L\rightarrow\infty$ does the non-locality vanish.  Physically, this corresponds to completely separating the doubled regions of the metric off to opposing infinities.  No entanglement\cite{raamsdonk} can arise in such a spacetime as the regions have each receded to infinities but in opposing directions.

Consequently, when the full brane structure of IIB string theory is considered,  an alternative to the standard geometric interpretation of the entanglement entropy must be constructed.  In the full 10-dimensional structure, some singularities can be circumvented.  Hence, we conjecture that the entanglement entropy should be constructed from the drawing the minimal mass (a type of current) in 10-dimensional spacetime.  The area of this surface we submit will be the true entanglement entropy.  Note the projection of this surface to AdS$_5$ does not preserve minimality because of the presence of curvature.   We are advocating more than just an extension of the geometric interpretation of the entropy to AdS$_5\times X$, where $X$ is a compact Einstein manifold, as has been done recently\cite{smooth}.  What is required here is a generalization because singularities appear explicitly in the bulk. 

\section{Closing Remarks}

We have shown here that the full structure of IIB string theory is needed to remove the non-localities that arise in boundary conformal theories that border hyperbolic spaces. What this work ultimately tells us is that the gauge-gravity duality as a statement about strictly hyperbolic spacetimes with complete metrics is not a theory about local conformal theories.  The boundary theories contain fractional conformal Laplacians and hence are non-local.  Consequently, the standard implementation of the gauge-gravity duality, in which mechanisms such D3-branes leading to metric incompleteness are absent, must yield local CFTs on the boundary.  Metrics underlying the Randall-Sundrum\cite{lrandall} work are candidates for removing the non-localities.

Relatedly,  all examples in which the gauge-gravity correspondence has been worked out explicitly (and asymptotically explicit is included here), either D3-branes  (which we have shown remove the boundary non-locality) are explicitly included in the bulk\cite{maldacena} or D-3 branes are absent and the boundary theory contains explicitly non-local operators\cite{SYK,SYK1}.   
On some level, this is not surprising because at the core of gravity are the equivalence principles which preclude local observables.  As a result, any  theory with gravity necessarily has less observables than a theory without it.  Consequently, an {\it a priori} correspondence between a bulk theory of gravity and a local boundary CFT must include some added features in the bulk that would ultimately permit a purely local theory to emerge on the boundary.

Since there is no guarantee that the current-carrying degrees of freedom in strongly correlated electron matter have a local description, the standard implementation of the gauge-gravity correspondence without the inclusion of D3-branes ultimately has utility.  The non-local  interactions that arise in this case can be useful in describing fractional gauge fields in strongly correlated quantum matter as in the strange metal of the cuprates\cite{anomdim} or yield a method to obtain unparticle propagators\cite{Georgi2007a,gabadadze}.   In fact, the Higgs mechanism we have proposed here provides a general way of engineering boundary propagators with arbitrary anomalous dimensions.  The precise form of the entanglement entropy in IIB string theory remains purely conjectural as of this writing.

\section{Appendix}

Here we review some of the basics of the correspondence. For simplicity of notation, we consider the case $d=4$.
We fix $AdS_5$ with a given metric $g_{\mu \nu} dx^\mu dx^\nu$ with fixed conformal infinity, which we take to be the conformal class of the round sphere $S^4$. Of course, if we insist on $g_{\mu \nu} dx^\mu dx^\nu$ being Einstein, this uniquely determines it as the classical AdS$_5$ (this is still true if the conformal class is sufficiently close to the round one\cite{grahamlee}).
Let $S$ be an effective action in the bulk. For instance this could be of the form,
\beq
S=S(g_{\mu \nu}, A_\mu, \phi, \cdots).
\eeq

\noindent
We let $L$ be the Lagrangian of the boundary CFT. The primary operators at the boundary specify the spectrum of the said CFT.
The correspondence dictates that one associates an operator $\cal O$ at the boundary to a field $\phi$ in the bulk. The operator is associated with the source $\phi_0$ by
\beq
L_{CFT} + \int d^4x \phi_0 \cal O,
\eeq
which determines the partition function to be
\beq
\langle e^{\int_{S^d}\phi_0{\cal O}}\rangle_{\rm CFT}=Z_S(\phi),
\eeq
where $S$ is the given theory in the bulk evaluated on shell, so $\phi$ is an extension of $\phi _0$ satisfying the classical equations of motion.
In order to calculate the (connected) n-point functions of the boundary theory, we write
\beq 
e^{W(\phi)}= \langle e^{\int_{S^d}\phi_0{\cal O}}\rangle_{\rm CFT},
\eeq
and then calculate
\beq
 \langle {\cal O} \cdots {\cal O}\rangle _c=\frac{\delta ^n W}{\delta \phi _0^n}\mid _{\phi _0=0}.
\eeq
We now specialize to the case where $S= \sqrt{g} \left( g^{\mu \nu} \partial _\mu \phi \partial _\nu \phi +m^2 \phi^2\right)$, the Klein-Gordon action.
Since we are meant to calculate $S(\phi)$ on shell, by integration by parts (eq. \eqref{intbyparts}), we find that the finite part of $S(\phi)$ is
\beq 
{\rm pf} S(\phi)=-d\int_{y=0} \; \phi _0\,g_0,
\eeq
where we expand the classical solution as $\phi =F y ^{d-s} + G y^s,\quad F,G\in\mathcal C^\infty(\mathbb H),\quad F=\phi_0+ O(y^2), \quad G= g_0 + O(y^2)$ 
wehere $g_0= (-\Delta )^\gamma \phi _0$ as we demonstrate in the text.
Therefore this determines $W$ and shows that there is no  ~n-point function for $n\neq 2$. 
This computation holds also for any $g_{\mu \nu}$ which is conformally compact, thus indicating that we have exactly determined the dual of the Klein-Gordon theory. 
To recover the full Maldacena duality one needs to add the Dd-brane constructions discussed in the text or perhaps other features of Type IIB string theory.   As we demonstrate the non-localities vanish as $N\to +\infty$.  


\textbf{Acknowledgements} 
We thank G. Vanacore for carefully reading an earlier draft and the NSF DMR-1461952 for partial funding of this project and the J. S. Guggenheim Foundation for providing a fellowship to P. W. P.

%
\end{document}